# Willingness to Pay for Community-Based Health Insurance among Rural Households of Southwest Ethiopia

Melaku Haile Likka[1†], Shimeles Ololo Sinkie[2] and Berhane Megerssa[2]

[1]Department of Medical Informatics; Graduate School of Integrated Arts and Sciences; Kochi University, Kochi, Japan.

[2]Department of Health Economics, Management, and Policy; Institute of Health Sciences; Jimma University, Jimma, Ethiopia.

[†]Corresponding author

Email addresses:

MHL: likkamelaku@gmail.com

SOS: shimelessos@yahoo.com

BM: berhanemegerssa2004@gmail.com

# Abstract

Use of healthcare services is inadequate in Ethiopia in spite of high burden of diseases. User-fee charges are the most important factor for this deficiency in healthcare utilization. Hence, the country is introducing community based and social health insurances since 2010 to tackle such problems. This study was conducted cross-sectionally, in March 2013, to assess willingness of rural households to pay for community based health insurance in Debub Bench district of Southwest Ethiopia.

Two stage sampling technique was used to select 845 households. Selected households were contacted using simple random sampling technique. Double bounded dichotomous choice method was used to illicit the willingness to pay. Data were analyzed with STATA 11. Krinsky and Rob method was used to calculate the mean/median with 95% CI willingness to pay after the predictors have been estimated using Seemingly Unrelated Bivariate Probit Regression.

Eight hundred and eight (95.6%) of the sampled households were interviewed. Among them 629(77.8%) households were willing to join the proposed CBHI scheme. About 54% of the households in the district were willing to pay either the initial or second bids presented. On average, these households were willingness to pay was 162.61 Birr per household (8.9 US$) annually.

If the community based health insurance is rolled out in the district, about half of households will contribute 163 Birr (8.9 US$) annually. If the premium exceeds the amount specified, majority of the households would not join the scheme.

**Key words**: community based health insurance, willingness to pay, contingent valuation method, double bounded dichotomous choice, Krinsky and Robb, rural households, Ethiopia.

# Introduction

In terms of access to healthcare and various health indicators, Ethiopia positions lower even as compared to other less developed countries [1].The nation bears a high burden of ailments mainly due to avoidable diseases and conditions. In spite of this, utilization of modern healthcare services is inadequate with outpatient department attendance of 0.3 per capita per year in 2011[2]. User-fee charge is the most important reason for meagre healthcare utilization in Ethiopia [3].

Out-of-pocket charges covers majority of the private healthcare expenditure and 34% of the total healthcare expenditure in Ethiopia. In comparison with other low income countries, user-fee charge is higher in this country [4, 5]. Nevertheless, per capita public spending for health remains far below even the average of Sub-Saharan Africa and low income countries, with US$ 20.77 per year in 2010/11 fiscal year [4].

As healthcare expenses were catastrophic and had negative impacts on livelihoods to the majority of families in Ethiopia, alternative mechanisms such as health taxes and pre-payment schemes were suggested for financing healthcare expenditures [6, 7]. Consequently, the government is acting on demand side factors that limit access to healthcare. One of the actions is the introduction of health insurance schemes.

Two types of health insurance schemes are being introduced in Ethiopia since 2010; social health insurance (SHI), which is targeted to cover 10.46% of the population; and, community-based health insurance (CBHI), which is planned to cover about 84% of the population of the nation [8].

Community-based social dynamics & collective risk sharing, solidarity, participatory decision-making & administration, non-profitability & voluntary membership are characteristics of community-based health insurance schemes (CBHIS). The schemes provide free health care services, included in its benefit packages, for its individual or household members who enroll in it by paying premium required [9].

The benefit packages of community-based health insurance (CBHI) in Ethiopia is supposed to include all family healthcare facilities and curative cares that are included in the basic health packages of Ethiopia, when the scheme is scaled up to full operation [10].

Prior to installing CBHIS, its suitability within the community and its feasibility (sustainability) should be decided. Sustainability is determined by the scheme's design, while suitability must be tested in community surveys or in pilots through measurements of the people's willingness to join (WTJ) and pay (WTP) before fully implementing CBHI [11].  Generally, demand data are rarely produced or used to design health insurance schemes in developing countries [12].

In Ethiopia, even though CBHIS is just being introduced, a few studies were conducted to determine the acceptability of such pre-payment schemes.  In 2007, households in Jimma town, who were willing to join *iddir* based health insurance schemes, reported to pay 7.60 Ethiopian Birr (ETB) (0.89 US Dollar) monthly per household for membership, if it had been rolled out.  Monthly income, educational status and relation of respondent to household, participation in *iddirs* had statistically significant effect on willingness to pay for *iddir* based health insurance schemes (IBHIS) [13]. Another study conducted in rural Ethiopia  in 2001 revealed that the households pay 10 birr (US\$ 1.22) in cash, or 14 birr (US\$ 1.71 ) in kinds per month per household  which is equivalent to 120 birr (14.6 US\$ ) in cash or 168 birr (20.5 US\$) in kind per household per year [14]

In the current study area, there are no any published data on WTP.  Hence nothing is known about the amount the households are willing to pay for the CBHIS which is to be implemented nationally in near future.  This initiated the investigators to conduct this study. The objective of this study was to assess willingness to pay for community-based health insurance scheme among rural households in *Debub* Bench District of Bench *Maji* Zone, South-west Ethiopia, 2013.

In this article, the term "*kebele*" is used to refer for lowest administrative body in Ethiopia which comprises at least 1000 households or population of 5000 people.

## Methods and Materials

A community based cross-sectional study was conducted in rural *kebeles* of *Debub* Bench District, *Bench Maji* Zone, southwest Ethiopia in March 2013 to determine willingness to join [15]and pay for community-based health insurance scheme among rural households of the district. Debub Bench District is one of the 9 districts of Bench Maji Zone in Southern Nations, Nationalities and People's Regional State (SNNPRS). It is located 858 kilometers south west of *Hawassa*, the capital of SNNPRS, and 588 kilometers south west of Addis Ababa, the capital of Ethiopia.  The district has 25 rural *kebeles* and 1 town administrations. The population of the district was projected to be 127,477 in 2012 [16]

A sample size of 845 was calculated by a single proportion formula, taking P = 50% (expected rate of acceptance of the initial or second bid), considering a design effect of 2 and an anticipated non-response rate of 10%.

A two-stage sampling technique was used to select participating households. The primary sampling units were *kebeles*. From 25 rural *kebeles*, 8 had been selected randomly. The selected *kebeles* were further divided into villages (sub-*kebeles*). The sampled households were allocated to each *kebeles* proportionate to the size of households. The number of sampled households of each *kebeles* was further allocated to all sub-*kebeles*. Then the participating households were selected from all sub- *kebeles* of the 8 *kebeles* using simple random sampling techniques.

All residents whose age were 18 years and above and who lived for more than six months in the *kebeles*, and the heads or spouses not employed in formal sectors, were eligible for the study. The house numbers of the households were used to develop sampling frames of the households of each sub-*kebeles*.

Various methods have been developed to assess the value of public and non-market goods and services. Contingent valuation method (CVM) is one of the direct methods used to illicit individuals' monetary values for public and non-market goods or services [17]

Suppose that the indirect utility of an individual i depends on buying health insurance policy and income y. let $q^1$ and $q^0$ measure the level of utility associated with and without health insurance, respectively, WTP is the amount of money an individual is WTP as a premium, X represents the vector of other factors (such as age, sex, education, health status, etc.) that may affect the preferences of individuals, $\pi$ shows the perceived probability of falling sick and $\varepsilon$ captures other factors that are unobservable to the researcher. Then, the individuals will buy the health insurance policy only if $v[(q^1, y\text{-WTP}, X, \pi) + \varepsilon_1) \geq v[(q^0, y, X, \pi) + \varepsilon_0)$. [17].

Where: $\varepsilon_{1 \text{ and}} \varepsilon_0$ are random errors distributed independently with mean zero;

There are a number of methods which uses to illicit WTP. One of such methods is double bound dichotomous choice contingent valuation method (DBDC-CVM) which was employed in this study. This elicitation method was used since it is information intensive, asymptotically more efficient, less sensitive to starting point and strategic biases, and reduces the demand for a large sample size [17].

In DBDC-CVM, each respondent is asked if s/he is willing to pay the first bid. If s/he says 'yes' to the first bid, a second higher bid is given and her/his willingness to pay is asked. If s/he says 'no' to the initial bid, a second lower bid is provided. If s/he says 'no' or 'yes' to both the first and the second bids then s/he is asked to the mention maximum amount of money that s/he is willing to pay [17]

A pre-tested structured questionnaire was used which comprised of variables on demographic and socioeconomic characteristics, health status and health care and health related variables, social capital, participation in *iddirs,* willingness to participate in community based health insurance schemes and debriefing questions. (*Iddirs* are funeral associations in Ethiopia that ensure a payout in cash and in kind at the time of a funeral for a deceased member of the family of a member of the group [17]. The data collection instrument was modified based on the findings of the pre-test before it was administered to the respondents.

In this study, six different starting bids were used to reduce starting bid bias associated with CVM. The initial bids have been assigned to be 500, 400, 300, 180, 100 and 50 Birr (which were equivalent to 27.3, 21.9, 16.4, 9.8, 5.5 and 2.7 US$ respectively as converted according to the exchange rate: **1US**$=18.3 Birr, at the time of data collection) randomly. The six initial bids were derived from premium of the pilot phase of Ethiopian CBHI, and other literatures conducted in Ethiopia on WTP for health insurances, based on calculations on per capita health expenditure of Ethiopians, etc.

After the interviewers explained the CBHI scenario to the respondents who were willing to join the scheme, they selected one of the six initial bids through lottery method. Then the respondent was asked his/her willingness to pay as annual premium of CBHIS for the household.

Data were collected by eight first cycle school teachers. Two health professionals were recruited as field supervisors. Training was provided to both the data collectors and supervisors for two days and one day respectively.Then, the data collected were entered and validated in double entry validation of Epidata 3.0and exported to STATA 11.

Both univariate and multivariate techniques to analyse the nature of the explanatory variables were included in the model. A frequency distribution of maximum price willing to pay was run and relative demand at different prices (i.e., the percentage of respondents who would pay each price for the scheme) was estimated.

In DBDC – CVM specification, the $WTP_i$ value is not directly observed. However, we observe a range of WTP values from the survey response.The WTP lies somewhere (a) between the two bids ('yes'–'no', 'no'–'yes'), (b) below the second bid ('no-no') or (c) above it ('yes'–'yes'). This implies that neither a continuous nor a simple dichotomous model can be used to analyze the response of households [18]. Therefore, bivariate probit regression was used to identify factors affecting the WTP of households for CBHIS.

Since the second response was related to the information provided in the first response, estimating the two responses independently or pooling them together to estimate a single equation may give invalid results [18]. Therefore, seemingly unrelated bivariate probit  regression ( SUBPR) model, which fits to two dichotomous responses and takes the possible dependence of the second response to the information provided in the first response, was used to identify factors affecting the WTP of households for CBHIS. The fitness and the overall significance of explanatory variables were examined by using the z-test. To control the potential biases from non-normality, robust estimators were used. This form of regression is also helpful to reduce the effects of heteroscedasticity.  A conditional marginal effect (partial effect) or discrete changes for dummy variables were calculated to measure the effect of each independent variable on the probability of accepting the initial and second bids, holding all other independent variables constant at their mean.

In the model two sets of parameter estimates are available from double bounded question. In this case it must be decided which estimates to use to calculate mean WTP measure. So, parameter estimates from the first equation (initial bid) were generally used in the computation of mean willingness to pay because the second equation parameters are likely to contain more noise in terms of anchoring bias where the respondent is assumed to take the cue from the first bid while forming his WTP for the second question.

Krinsky and Robb statistics for Confidence Intervals was used to calculate the mean WTP. Parameter estimates from the first equation (initial bid) were generally used in the computation of mean willingness to pay because the second equation parameters are likely to contain more noise in terms of anchoring bias where the respondent is assumed to take the cue from the first bid while forming his WTP for the second question.

Ethical clearance was obtained from Ethical Review Committee of Public Health and Medical Sciences Faculty of Jimma University. Support letters were obtained from *Debub Bench* District administration. Permission was sought from the *Kebeles* administration before conducting the study.

## Results

Of 845 sampled households, 808 participated in the study which yielded a response rate of 95.6%. The median age of respondents was 32, ranging 18-87 years. Among them, 574 (71 %) were male, 550 (68.1%) Protestant Christians, 519 (64.2%) Bench ethnic groups, 675 (83.5%) married (574 monogamous/monandrous and 101 polygamous), 615(76.1%) farmers, 388 (48%) had no education and 654 (80.9%) were head of the household. The average number of the household members was 5.4.The median annual household income, as estimated from the amount earned from sales of coffee, khat, maize, cassava, and other local products such as fruits, honey dairy products, etc., in one year time, was 2475 ETB, ranging between 100- 1860 ETB. From the analysis of the wealth index, 31.6% of the households were found in the second and 24.0% in the highest wealth quintiles (table 1).

Table 1: Demographic, socioeconomic, and health and health related characteristics of the study participants in Debub Bench district, southwest Ethiopia, 2013

| Description | | Frequency (%) |
|---|---|---|
| **Demographic and socioeconomic characteristics** | | |
| Sex of the respondent (n=808) | Female | 234(29.0) |
| | Male | 574(71.0) |
| Relationship (n=808) | Head | 654 (80.9) |
| | Spouse | 132 (16.3) |
| | Others(Child, parent) | 22 (2.7) |
| Religion of the respondent(n=808) | Protestant | 550 (68.1) |
| | Orthodox | 199(24.6) |
| | Muslim | 41 (5.1) |
| | Others | 18 (2.2) |
| Marital status of the respondent (n=808) | Monogamous/monandrous | 573 (70.9) |
| | Polygamous/polyandrous | 101 (12.5) |
| | Single | 55 (6.8) |
| | Widowed | 24(3.0) |
| | Divorced | 55(6.8) |
| Occupation of the respondent (n=808) | Farmer | 615(76.1) |
| | Housewife | 113(14.0) |
| | Merchant | 36(4.5) |
| | Student | 30(3.7) |
| | Others | 14(1.7) |
| Ethnicity of the respondent (n=808) | Bench | 519(64.2) |
| | Amhara | 116(14.4) |
| | Kaffa | 64 (7.9) |
| | Others | 109(13.5) |
| Educational status of the respondent (n=808) | Illiterate | 388(48.0) |
| | Read and write | 221(27.4) |
| | Grade 1-8 | 178(22.0) |
| | >=Secondary school | 21(2.6) |
| wealth quintile of the household (n=808) | Lowest wealth quintile (index< 20th percentile) | 70 (8.7) |
| | Second wealth quintile (index 20th-40th percentile) | 255 (31.6) |
| | Middle wealth quintile (index 40th-60th percentile) | 164(20.3) |
| | Fourth wealth quintile (index 60th-80th percentile) | 125 (15.5) |
| | Highest wealth quintile (index>80th percentile) | 194 (24.0) |
| | Lower than 1100 birr | 198(24.5) |
| Category of annual income (n=808) | 1100-4300 birr | 409 (50.6) |
| | More than 4300 birr | 201 (24.9) |
| **Health and health related variables** | | |
| Self-reported health status of the household (n=808) | Very poor | 50 (6.2) |
| | Poor | 166 (20.5) |
| | Medium | 270 (33.4) |
| | High | 224 (27.7) |
| | Very high | 98 (12.1) |
| Persons with chronic illness &/or disability in the HH(n=808) | No | 747 (92.5) |
| | Yes | 61 (7.5) |
| Any illness encountered during the past 3mths (n=808) | No | 558 (69.1) |
| | Yes | 250 (30.9) |
| Seek of medical treatment for the recent episode (n=250) | No | 19 (7.6) |
| | Yes | 231 (92.4) |
| Get treatment (n=231) | No | 12 (5.2) |
| | Yes | 219 (94.8) |
| place of treatment (n=219) | Private Heath Facility | 90 (41.1) |
| | Public health center | 65 (29.7) |
| | Public hospital | 49 (22.4) |
| | Other (self-treatment, traditional healer and local drug vendor | 15 (6.8) |
| Reasons for going there (n=219) | The HF was physically accessible | 104 (47.5) |
| | The HF was not expensive | 18 (8.2) |
| | The health facility not too crowded | 19 (8.7) |
| | The health service was effective | 66 (30.1) |
| | Other (specify) | 12 (5.5) |
| Reasons for not getting treatment (n=12) | No enough money | 9 (75.0) |
| | Others (too far, self-limiting) | 3 (25.0) |
| Coverage of the health care cost (n=219) | Self | 204 (93.2) |
| | Others (free, community) | 15 (6.8) |
| Satisfaction with health care service and costs (n=219) | very dissatisfied | 23 (10.5) |
| | Dissatisfied | 61 (27.9) |
| | Neutral | 8 (3.7) |
| | Satisfied | 111 (50.7) |
| | Very satisfied | 16 (7.3) |
| Perceived quality of the health care service in the district (n=219) | very low | 20 (9.1) |
| | Low | 76 (34.7) |
| | Neutral | 24 (11.0) |
| | High | 87 (39.7) |
| | Very high | 12 (5.5) |
| Concern of the household for covering health care costs (n=219) | Very difficult | 77 (35.2) |
| | Difficult | 110 (50.2) |
| | Not difficult | 32 (14.6) |
| Means of getting money for health care payment (n=187) | Drew from the savings | 38 (20.3) |
| | Borrow from someone | 27 (14.4) |
| | Assisted by relatives | 68 (36.4) |
| | Undertaken extra work | 2 (1.1) |
| | Sell capital assets such as cows | 33 (17.6) |
| | Cut back on other things, food, etc | 19 (10.2) |
| Borrow money for medical costs within last year (n=808) | No | 530 (65.6) |
| | Yes | 278 (34.4) |
| The nearest conventional health institution to the respondents' home (n=808) | Health center | 373 (46.2) |
| | Clinic (Private) | 367 (45.4) |
| | Hospital (Gov) | 68 (8.4) |

Seven hundred and forty seven (92.5%) of the households were participating in *iddirs*. Out of them 635 (85%) of the households were participating in one *iddir*) and the remaining 112 households in more than one *iddirs*. The median contribution of the households to *iddirs* was 1 ETB per month (0.055 US$) with range of 1-4 ETB (0.055-0.22 US$).

Individual level social capital of the community was also assessed. Two hundred and thirty three (28.8%), 545 (56.4%) and 119(14.7%) of the households were of low (lower than the $25^{th}$ percentile), middle (between $25^{th}$ and $75^{th}$ percentiles) and high (above $75^{th}$ percentiles of horizontal trust index) individual level horizontal trust respectively. Also, 222(27.5%), 451(55.8%) and 135 (16.7%) of the households were of low, middle and high individual level reciprocity respectively.

With respect to health status and health related variables, 50 (6.2%) of the respondents evaluated their family's health status to be very poor and 98 (12.1%) very high. Sixty one (7.5%) of the participants had at least one member with chronic disease or disability; and 250 (30.9%) of the households had at least one member who had encountered illnesses 3 months prior to data collection. Among the ill 231 (92.4%) had sought treatment for the illnesses they experienced, and 219 (94.8%) got treatment. The remaining 12 did not get treatment because of, mainly lack of money.

Of 219 who got treatment, 41.1% preferred to go to private clinics. They preferred the specified institutions because of its physical accessibility (47.5%), effective service (30.1%), not too crowded (8.7%), not expensive services (8.2%), or other reasons (5.5%).

The median health expenditure of the 219 households which sought treatments was 170 ETB (10.4 US$) with range of 18 to 2000 ETB (1-109.3 US$). Two hundred and four (93.2%) of the households covered the medical expenses by themselves. One hundred and eighty seven (85.4%) of these households reported that it was (very) difficult to cover payments for treatments. As a result, 68 (36.4%) of them were assisted by relatives to cover the medical costs; 38 (20.3%) drew from their savings, and 27 (14.4%) borrowed from someone. The remaining had to sell capital assets such as cows (17.6%), cut back on other things, food, drink, cloth etc. (9.1%), undertook extra works and search for other means(2.2%) to cover the payments for treatment.

Of 808 respondents, 278 (34.4%) households reported that they had borrowed money for covering health care expenses within one year before the data were collected. The median amount that these households borrowed was 200 ETB (10.9US$), ranging 30-2000 ETB (1.6-109.3 US$).

Regarding the distance of home of the household to the nearby health facility (private clinics, health centre or public hospital), it was reported that the median time it takes to reach the nearby health facility was 50 minutes, range between 3 minutes to 180 minutes(table 1).

Among the participants, 629 (77.8%) were willing to join the proposed community based health From these 629 households, 54 (8.6%) were willing to pay both the first and second bids (replied "yes"-"yes" for both first and second higher bids); 204 (32.4%) accepted the initial bid but refused the second higher bid ("yes"-"no"); 137 (21.8%) refused the initial but accepted the second lower bids ("no"-"yes") and 234 (37.2%) responded "no" for both the initial and the second lower bids ("no-no").

This implies that 54.2% of the households were willing to accept either the first or the second bids. The mean amount of initial bid presented and accepted was 224.18 birr (12.25 US$) and 149.42 birr (8.17 US$) respectively. Similarly, the mean amount of the second bid presented and accepted was 204.02 birr (11.15 US$) and 121.44 birr (6.64 US$) respectively.

Of 234 individuals who rejected both bids, 211 (90.17%) were willing to pay some amount of money (average of 48.8 birr (2.7 US$)) which is less than the second lower bid presented. The remaining 23 (3.7% of the total respondents) reported 0 WTP.

The seemingly unrelated bivariate probit model indicates that the model fitness is significant (Wald $\chi 2$= 285.37, Prob> $\chi 2$= 0.0000). The likelihood ratio test for $\rho = 0$ is significant (i.e, the null hypothesis that the correlation of error terms in two equations is zero is rejected) indicating endogeneity of the two equations.

The predicted probability of accepting the initial bid was 31.4% and that of the second bid was 18.2% at mean of the independent variables. The predicted probability that the initial bid is accepted was 17.7% at mean, given the second bid is accepted and that of the second bid, given the initial bid is accepted was 10.24% at the mean of the predictor variables.

The bid amounts were negatively and significantly associated with WTP. The marginal effects at mean of the variables show that the probability of accepting the initial and second bid reduces by 0.3% and 0.2%, respectively for every unit increment of the bids, other variables kept constant at their mean (table 2).

Demographic factors such as age, sex, relationship of the respondent to the household and family size have no significant effect in accepting both presented bids. Other religious group members were about 30% less likely to accept the initial bid than Protestants, ceteris paribus. Other categories of religion were not significantly different to Protestants in accepting the second bid (table 2).Furthermore, singles and widowers were about 26% less likely to accept the initial bid than the monogamous but widowers were 48% more likely to accept the second bid, ceteris paribus. Housewives were 23.2% and 10.2% less likely to accept the initial and second bids respectively than farmers (table 2)  Ethnically, *Amhara* ethnic groups were 20.3% and 14.2 % more likely to accept the initial and second bids respectively than Bench ethnic groups (table 2)

Educational status did not significantly affect the probability of accepting the second bid. But respondents who completed grade 1-8 were 12.3 % less likely to accept the initial bid, ceteris paribus (table 2)

Annual income of the households is positively and significantly associated with WTP. Households whose annual income is < 1100 birr were 25 and 11 percent less likely to accept the initial and second bids respectively than those whose income is 1100-4300, keeping other factors constant at mean. But those whose income is more than 4300 birr were 40.7 and 14.5 percent more likely to accept the bids respectively than those whose income is 1100-4300 birr. Wealth quintiles of the households had no significant effects in probability of accepting the second bids. But households at fourth and highest wealth quintiles were more likely to accept the first bid than those in the second quintile, ceteris paribus.

Individual level horizontal trust and reciprocity were not significantly associated with the probability of accepting the initial bid. But individual level reciprocity was positively associated with the probability of accepting the second bid. Contrary to this, community level reciprocity is negatively associated with the probability of accepting both bids. Ceteris paribus, communities with higher community level reciprocity were 14% and 9 % lower in the probability of accepting the initial bid and the second bids, respectively than those in lower reciprocity (table 2).

Participation in *iddirs*, number of *iddirs* the households participate in and the amount the households contribute for *iddirs* were not significantly associated with probability of accepting both bids (table 2).

The health status of the household was not significantly associated with accepting both bids. The presence of household members with chronic diseases or disability, getting treatment for recently occurred illness in the household and borrowing money for health care payments had no significance on the probability of accepting the initial bid. But households which borrowed money for covering medical treatment within one prior to the study period were 10% less likely to accept the second bid than those who did not borrow. And households which have at least one member with chronic disease or disability were 23.7 % more likely to accept the second bids than those with no such condition, ceteris paribus.

In addition households which did not get treatment for recently occurred ailments had 42.8% increased probability to accept the second bid than who did not. Perceived quality of healthcare system of the district, borrowing money for treatment and the amount of money borrowed, were significantly and positively associated with WTP (accepting both the bids).

Table 2: Factors affecting the probability of accepting the presented bids, Debub Bench District, southwest Ethiopia, 2013

| Seemingly unrelated bivariate probit | | | Number of obs = 629 | | |
|---|---|---|---|---|---|
| Log pseudo likelihood = -481.964 | | | Wald χ2 (104) = 402.41 | | |
| | | | Prob> χ2= 0.0000 | | |
| **INITIAL BID ACCEPTED (0-no, 1-yes)** | | | **2nd BID ACCEPTED (0-no, 1-yes)** | | |
| Coef. | Robust St.Err. | dy/dx | Coef. | Robust Std.Error | dy/dx |
| Initial Bid | -.0078689* .0007218 | .0027916 | ------------ | --------- | ------- |
| Second Bid | ----------------------- | ------ | .0073114* | .0010047 | -.0019308 |
| **RELIGION OF THE RESPONDENT (PROTESTANT= REFERENCE)** | | | | | |
| Others | -1.523709* .5016404 | -.3019703 | | | |
| **MARITAL STATUS OF THE RESPONDENT(MONOGAMOUS= REFERENCE)** | | | | | |
| Single | -1.107087* .5341466 | -.2689274 | | | |
| Widowers | -1.01149* .4491066 | -.2535331 | 1.341941* | .4259323 | .4815235 |
| **OCCUPATION (FARMERS- REFERENCE)** | | | | | |
| Housewives | -.790117** .3066464 | -.2318082 | -.4528139** | .2656793 | -.1018046 |
| Students | 2.620906* .5396892 | .6875264 | | | |
| **ETHNICITY OF THE RESPONDENT (BENCH-REFERENCE)** | | | | | |
| Amhara | .535567** .2927219 | .202547 | .4705087** | .2597543 | .1422649 |
| **EDUCATIONAL STATUS (NO EDUCATION-REFERENCE)** | | | | | |
| Grade 1-8 | -.3715163** .2124057 | -.1235618 | | | |

**COMMUNITY LEVEL SOCIAL CAPITAL (LOW-REFERENCE)**

| | | | | | | |
|---|---|---|---|---|---|---|
| Horiz trust | .8077952* | .1807314 | .3032166 | | | |
| Reciprocity | -.4152626* | .1870592 | -.1471779 | -.347788* | .1608179 | -.0924259 |
| Chronic(no-ref) | | | | .7269238* | .2385108 | .2373678 |
| Illness(no-ref) | | | | -.7576756** | .3667883 | -.1804526 |
| Get tr(yes-ref | .1733251** | .4470481 | .0623355 | 1.393746* | .4906148 | .4280534 |
| **HEALTHCARE COST (0 BIRR- REFERENCE)** | | | | | | |
| <1-100 | .7890152* | .2514537 | .3034718 | | | |
| Quality of HC | .220766** | .1124895 | .0783194 | .1931433** | .1013 | .0510069 |
| Borrow for trx (yes-ref) | | | | -.3939742* | .1669909 | -.1000291 |
| Amount borrowed | .0015533* | .0003469 | .000551 | .0009709* | .0002873 | .0002564 |
| **NEAREST HEALTH FACILITY (HEALTH CENTRE-REFERENCE)** | | | | | | |
| Clinic | .8376163* | .1831483 | .2941749 | | | |
| Hospital | -1.982595* | .3872115 | -.3658371 | | | |
| TIME TO HF | -.0003122** | .0028168 | -.0001108 | | | |
| **WEALTH QUINTILE (SECOND QUINTILE-REFERENCE)** | | | | | | |
| Fourth quintile | .5294682* | .2466635 | .1998714 | | | |
| Highest quintile | .6621796* | .2139222 | .2455529 | | | |
| **ANNUAL INCOME (1100-4300 BIRR-REFERENCE)** | | | | | | |
| <1100 birr | -.8207394* | .2195967 | -.2490358 | -.4832123* | .1769152 | -.1114412 |
| >4300 birr | 1.098838* | .1910727 | .4072538 | .5005494* | .166892 | .1452105 |
| _cons | .2358282 | .8588142 | | -.3987128 | .8568627 | |
| /athrho | -.292473 | .1264636 | | -.292473 | .1264636 | |
| Rho | -.2844093 | .1162342 | -.2844093 | | .1162342 | |

Likelihood-ratio test of rho=0: $\chi2$ (1) = 5.3486  Prob> $\chi2$= 0.0207

NOTE: **- significant at p<0.10, *- significant at p<0.05

Health care cost the households incurred within 3 months prior to the study period was also associated with the probability of accepting the initial bid. In addition, households nearby to private clinics were 29.4% more likely to accept the initial bid than those nearby to health centers. But those nearby to hospital were 36.6% less likely to accept the initial bid than those nearby to health centers, ceteris paribus.

Distance of household to health facilities was negatively associated the probability of accepting the initial bid. As the time taken to reach the nearby health facility increases the probability of accepting the initial bid significantly reduced slightly (table 2)

**Mean/Median Willingness to Pay for Community Based Health Insurance and Probability of Accepting the Bids at Different Amounts**

The mean/median WTP, calculated based on the initial bids presented, was 162.61 ETB (8.9 US$) per year per household (95% CI: 142.28ETB (7.8US$) - 181.56 ETB (9.9 US$)). If the mean/median WTP were estimated based on the second bid, the amount would be 79.79 birr per year per household (95% CI: 44.48- 102.31 Birr). If the covariates were not considered, the Krinsky &Robb mean/median WTP would 149.45 birr (95% CI: [116.43, 177.71]).

Table3: The mean/median willingness to pay (in birrs), using Krinsky and Robb 95%confidence interval for the scheme in Debub Bench, Southwest Ethiopia, 2013

| Mean/median WTP (95 % CI) | ASL* | CI/MEAN | Equations used |
|---|---|---|---|
| 162.61(142.28, 181.56) | 0.0000 | 0.24 | initial bid |
| 79.79 (44.48, 102.31) | 0.0006 | 0.72 | second bid |
| 149.45(116.43, 177.71) | 0.0000 | 0.41 | Without covariates |

*: Achieved Significance Level for testing H0: WTP<=0 vs. H1: WTP>0
CI- confidence interval
WTP- willingness to pay

The mean WTP was estimated based on the first equation. Therefore, the mean WTP for the scheme in the study area is 162.61 ETB (95% CI: [142.28, 181.56]) per year per household (table 3). Most of the households (37.8%) who reported positive WTP preferred to pay the premiums quarterly a year flat rates (table 4).

Table 4: Preferred frequency of contribution of the premiums by the households in Debub Bench District, Bench Maji Zone, southwest Ethiopia (n=606)

| Preference | Frequency (%) |
|---|---|
| Annual flat rate | 106 (17.5) |
| Bi-annual flat-rate | 137 (22.6) |
| Quarterly a year flat-rate | 229 *(37.8) |
| Monthly | 134 (22.1) |

The probability of accepting bids at different amounts was predicted. From table 5, it is predicted that 50% of the households would be willing to pay the bid at the calculated mean WTP (162.61 ETB per year per households). Similarly, about 54% of the households would WTP if the bid is 150 birr; and the probability of accepting the bid would be 0 if its amount is more than 600 birr/year/annum (table 5).

Table 5: The probability of accepting the bids by the households at different bid amounts in Debub Bench Disrict, Bench Maji Zone, southwest Ethiopia.

| Amount of the bid in Birr | Predicted probability of accepting the bid | Amount of the bid in Birr | Predicted probability of acceptance |
|---|---|---|---|
| 15 | .877288 | 180 | .44557902 |
| 25 | .86056174 | 200 | .38429573 |
| 50 | . 81222441 | 250 | .24583195 |
| 75 | .75471409 | 300 | .1398237 |
| 79.79 | .74270503 | 360 | .06018193 |
| 100 | .68887928 | 400 | .03088064 |
| 125 | .61636688 | 450 | .01186562 |
| 150 | .53952184 | 500 | .0039666 |
| 162.61 | .50000073 | 600 | .00028894 |

As discussed earlier, the WTJ of the households for CBHIS in the study area is 77.8%. If the premium is set to be 162.61 ETB per annum per household, half of those who are WTJ the scheme would be able to pay. Therefore, the scheme would be accepted by only 38.9% of the households of the district at the specified premium. If the district has 23607 households (127,477 populations divided by 5.4 individuals per household), only 9183 of them would be members of CBHIS and they would contribute 1493175 birr per annum as membership payments if the scheme is rolled out by setting 162.61 birr per annum per household.

## Discussion

In this study, it was revealed that 54.2% of the households in the study area were willing to pay either the first or second bids presented. This figure is lower than that found in a study conducted from samples of four regions of Ethiopia in which 60% of the households were WTP the first or second bids [14]. The disagreement may be attributed to difference in the demand for pre-paid schemes among the sampled households in both studies. The current willingness of the households to join the proposed health insurance was lower (77.8%) than the previous one (94.7%). Another possible reason for lower willingness to accept the proposed bid may be differences in the study participants. The scope of this study is narrower in comparison with the existing one in which the participants were selected from four regions. It could also be suggested that the chronological gap between the study times may contribute for discrepancy. Future studies may unfold the reason for differences in WTP in these studies.

The current study indicated that the households were WTP about 163 birr (8.9 USD) annually per household whose family size, on average was 5.4.In a study conducted in rural parts of Ethiopia in 2004 that rural households willing to pay 10 birr (US$ 1.22)in cash, or 14 birrs (US$ 1.71) in kinds per month per household which is equivalent to 120 birr (14.6 US$) in cash or 168 birr (20.5 US$) in kind per household per year , with average family size of 6.6 per household[14]. Unfortunately, contribution of the households in kind has not been elicited in the current study. So, the comparison can be made between the amounts the households were WTP in cash in both studies. On the other hands, because of the devaluation undertaken by the government 2011 and deflation of ETB in relation to USD, the real and nominal amounts can be compared at different angles.

In terms of nominal contributions, as expected, the current WTP (163 birr per households per annum) is larger than that of existing one (120 birr per household per annum). The increment is reasonable as the real purchasing power of the national currency is being reduced time to time. But the WTP of the current study area is almost equal to that of the previous one which was elicited to be contributed in kinds. Hence, it is gives sense to say if similar alternative WTP were used in the current study area, the WTP could be greater than 163 birr per household per annum.

When the real values of WTP of both cases are compared, currently the people are less likely to contribute for community based health insurance schemes than that was 10 years earlier, while it was expected to increase. The possible reason for the decrement may be differences in the target population. But, more probably, family size may be the most important reason for the decrement in WTP for CBHIS between the populations studied in these studies. The mean family size in the current study was 5.4 per households but it is not statistically significant factor for WTP in the current study area; whereas, that of the households in the previous study was 6.4 per household and it was positive predictor in their decision to pay for CBHIS. Since CBHIS benefit package includes individuals of the member households, households with larger family size to tend to pay grater amount than their counterparts with lesser family size.

Another study conducted in Jimma town in 2009 revealed that the households were WTP 91.2 birr (10.68 USD) per annum for *iddir* based health insurance scheme [13]. This WTP is still larger than the current one. In Ethiopia, it is clear that ability to purchase and demand for health care and other goods and services significantly varies in urban to rural. This may be the sole reason for lesser WTP for CBHIS in the rural district studied. Another difference between these two findings is that majority of the respondents in previous study preferred to contribute monthly; whereas, in the current study area they prefer to contribute quarterly or annually. This may be because of differences in employment status of the populations. Majority of the population in urban areas are engaged in formal sectors; whereas rural populations are engaged in informal sectors.

Th fifth National Health Account Survey conducted in 2010/11 disclosed that annual per capita health expenditure of Ethiopians is 20.77 USD. The proportion of the expenditure that the households cover is 34% [4].When we convert this figure to the household of 5.4 family size, the expenditure would be 112.16 USD. Thirty four percent this amount that the households have been covered would be 38.1 USD per household per annum. This amount is four times larger than the reported WTP in the study area. This puts sustainability of CBHIS in the study area in doubt.

In the current study demographic factors such as age, sex, relationship of the respondent to the household and ethnicity were not significantly associated with the probability of the accepting the presented bids at 95% confidence level so did these factors in literatures discussed earlier[13,14]. But this was not consistent with findings in a number of literatures mainly from West African and Asian countries [19–23].

On the other hand, educational status and family size of the households were found to statistically insignificant predictors in the decisions of the households to accept the presented bids. These variables were significant predictors in the studies discussed above [13, 14] and other findings from West Africa and Asia [19–23]. Concerning educational status, the discrepancy might have occurred because of the homogeneity of the respondents in the current study in their educational status. More than three quarters of the current study participants were either illiterate or only able to read and write.

Socioeconomic variables (wealth index and annual income) had positive associations with WTP in the study area, consistently with most of existing literatures [19–23]. This is in line with economic theories.

## Conclusion

When community-based health insurance scheme is initiated in the district, the maximum price set as a premium should not exceed the 163 ETB (8.9 US$) for more than half of the households in the district would not able to accept the premiums. If the households are provided ways of contributing to CBHIS other than in cash contributions, such as in kind contributions, they might contribute more than the amount specified.

Most of health and health related situations are significant predictors for WTP implying that adverse selection might be a problem when the scheme is started. Therefore mechanisms to control such controls, such as co-payment should be considered.

## Acknowledgements

The authors are grateful to acknowledge Jimma University for providing financial assistance for the study. We also like to acknowledge the study participants, schools in the selected kebeles, administration of Debub Bench District, Health and Education Office of the district, and administrations of the selected kebeles.

## Conflict of interest

We, the authors declare that the funding institution or the authors have no any conflict of interests.

## References

1. United Nations Development Programme (UNDP). Human Development Report 2011: Sustainability and Equity: A Better Future for All. United Nations Development Programme (UNDP). United Nations Development Programme (UNDP); 2011.

2. Federal Ministry of Health, Health Care Financing Team, Policy, Planning and Finance General Directorate. Fourth National Health Accounts, 2007/08. Addis Ababa, Ethiopia; April 2010.

3. Federal Democratic Republic of Ethiopia Ministry of Health. Health Sector Development Programme IV Annual Performance Report Ethiopian Fiscal Year 2003. Addis Ababa, Ethiopia; 2011.

4. Federal Ministry of Health. Ethiopia's Fifth National Health Accounts, 2010/2011. Addis Ababa, Ethiopia: Federal Democratic Republic of Ethiopia Ministry of Health, USAID Health Sector Financing Reform Project; 2010 Apr.

5. World Health Organization. World Health Statistics 2011. WHO: Geneva; 2011.

6. Barnett I, Tefera B. Poor Households' Experiences and Perception of User Fees for Healthcare: a mixed-method study from Ethiopia. 2010 Jun; Available from: www.younglives.org.uk

7. WHO. The World Health Report: Health Systems Financing: The Path To Universal Coverage. [Internet]. Geneva; 2010 [cited 2012 Apr 23]. Available from: http://www.who.int/whr/2010/en/index.html

8. Federal Democratic Republic of Ethiopia Ministry of Health. HSDP IV 2010/11 –2014/15. Federal Democratic Republic of Ethiopia Ministry of Health; 2010.

9. Carrin G, Waelkens M-P, Criel B. Community-based health insurance in developing countries: a study of its contribution to the performance of health financing systems. Tropical Medicine & International Health. 2005 Aug 1;10(8):799–811.

10. Ministry of Health Ethiopia, Planning and Programming Department. Health Insurance Strategy. Ministry of Health Ethiopia; 2008.

11. Eckhardt M, Forsberg BC, Wolf D, Crespo-Burgos A. Feasibility of community-based health insurance in rural tropical Ecuador. Rev PanamSaludPublica. 2011 Mar;29(3):177–84.

12. Arkin-Tenkorang D. Health insurance for the informal sector in Africa: Design features, risk protection and resource mobilization. World Bank's Human Development Network Discussion Paper.; 2001.

13. Shimeles O, Challi J, Yohannes H, Belaineh G. Indigenous Community Insurance (Iddirs) As An Alternative Health Care Financing In Jimma City, Southwest Ethiopia. Ethiop J Health Sci. 2009 Mar;19(1):53–60.

14. Asfaw A, Braun J. Can community health insurance schemes shield the poor against the downside health effects of economic reforms? The case of rural Ethiopia. Health Policy. 2004;70:97–108.

15. Haile M, Ololo S, Megersa B. Willingness to join community-based health insurance among rural households of Debub Bench District, Bench Maji Zone, Southwest Ethiopia. BMC Public Health. 2014;14(591).

16. Federal Democratic Republic of Ethiopia Population Census Commission: Summary and Statistical Report of the 2007 Population and Housing Census Results. Central Statistical Authority; 2008.

17. Abay Asfaw, Emily Gustafsson, Wright Jacques van der Gaag. Willingness to Pay for Health Insurance: An Analysis of the Potential Market for New Low Cost Health Insurance Products in Namibia. Amsterdam Insitute for International Development. 2008;RS 0801/2.

18. Cameron TA, Quiggin J. Estimation Using Contingent Valuation Data from a "Dichotomous Choice with Follow-Up" Questionnaire. Journal of Environmental Economics and Management. 1994;28:218–34.

19. Dror, M, Radermacher, R, Koren, R. Willingness to pay for health insurance among rural and poor persons: Filed evidence from seven micro health insurance units in India. Health Policy. 2007;82:12–27.

20. Ali Asgary, Ken Willis, Ali Akbar Taghvaei, MojtabaRafeian. Estimating rural households' willingness to pay for health insurance. Eur J Health Econom. 2004;5:209–15.

21. Curt Lofgren, Nguyen X Thanh, Nguyen TK Chuc, Anders Emmelin, Lars Lindholm. People's willingness to pay for health insurance in rural Vietnam Published: Cost Effectiveness and Resource Allocation. BioMed Central Ltd [Internet]. 2008 Aug 11;6(16). Available from: http://www.resource-allocation.com/content/6/1/16  2008 Lofgren et al;

22. Donfouet HPP, Essombè J-RE, Mahieu P-A, Malin E. Social Capital and Willingness-to-Pay for Community-Based Health Insurance in Rural Cameroon. Global Journal of Health Science. 2011 Apr 1;3(1):142.

23. OA Babatunde, TM Akande,, AG Salaudeen, SA Aderibigbe, OE Elegbede and, LM Ayodele. Willingness to Pay for Community Health Insurance and its Determinants among Household Heads in Rural Communities in North-Central Nigeria. International Review of Social Sciences and Humanitie. 2012;2(2):133–42.